\documentstyle[prb,aps]{revtex}
\begin{document}
\baselineskip20pt
\thispagestyle{empty}
\pagestyle{plain}
\title{
Weak antiferromagnetism due to Dzyaloshinskii-Moriya interaction 
in Ba$_3$Cu$_2$O$_4$Cl$_2$  }
\author{V.\ Yushankhai}
\address{ Max-Planck-Institut f\"ur Physik Komplexer Systeme, D-01187
Dresden, Germany \\
and Joint Institute for Nuclear Research, 141980 Dubna, Russia}
\author{M.\ Wolf, K.-H.\ M\"uller, R.\ Hayn, and H.\ Rosner}
\address{Institut f\"ur Festk\"orper- und Werkstofforschung (IFW) Dresden, 
D-01171 Dresden, Germany}
\date{\today}
\maketitle

\begin{abstract}
The antiferromagnetic insulating cuprate Ba$_3$Cu$_2$O$_4$Cl$_2$
contains folded CuO$_2$ chains with four magnetic copper ions ($S=1/2$) per
unit cell. An underlying multiorbital Hubbard model is formulated and the
superexchange theory is developed to derive an effective spin Hamiltonian for
this cuprate. The resulting spin Hamiltonian involves a   
Dzyaloshinskii-Moriya term and a more weak symmetric anisotropic exchange 
term besides the isotropic exchange interaction. The corresponding 
Dzyaloshinskii-Moriya vectors of each magnetic Cu-Cu bond
in the chain reveal a well defined spatial order. Both, the superexchange
theory and the complementary group theoretical consideration, lead to the same 
conclusion on the character of this order. The analysis of the ground-state
magnetic properties of the derived model leads to the prediction of an 
additional noncollinear modulation of the antiferromagnetic structure. This
weak antiferromagnetism is restricted to one of the Cu sublattices.  
\end{abstract}


\section{Introduction}

Electronic, superconducting and magnetic properties of the fast growing family
of copper oxides  
and oxychlorides have attracted much attention last years. The undoped parent
materials in the cuprate family are insulators showing a variety of
low-dimensional magnetic properties. This variety extends from the quasi
two-dimensional (2D) 
antiferromagnetic (AFM) behavior \cite{Johnston97,Greven95} with possible
admixture of a weak 
ferromagnetism \cite{Thio88,Ito97,Chou97,Kastner99,Eckert98} in the planar
compounds to quasi-1D magnetic 
properties, observable in the chain-like \cite{Johnston97,Matsuda97,Troyer97}
and spin-ladder \cite{Dagotto96} 
systems. The superexchange theory
\cite{Anderson59,Goodenough63,Moriya60,Geertsma80,Eskes93} provides the
necessary basis which 
allows to derive and to estimate the main interactions, including
anisotropic ones, responsible for the magnetic coupling of the copper
spins. The 
form, and especially the magnitude of several interaction constants of the 
resulting spin Hamiltonian, however, strongly depend on pecularities of the
Cu-O-Cu bond  configuration in different cuprates (see Ref.\
\onlinecite{Aharony98} and references therein). One interesting
class of magnetic cuprates contains competing magnetic subsystems, like for
instance tetragonal Ba$_2$Cu$_3$O$_4$Cl$_2$ built up of Cu$_3$O$_4$ planes
with two types of copper sites (CuI, CuII). \cite{Kipka1-76} 
It is well known that in Ba$_2$Cu$_3$O$_4$Cl$_2$  
(as well as in Sr$_2$Cu$_3$O$_4$Cl$_2$) the moments of the 
CuI and CuII atoms order antiferromagnetically at different temperatures
$T_{N,I}$ $\approx$ 330 $\ldots$ 380 K 
and $T_{N,II}$ $\approx$ 30 $\ldots$ 40 K, 
respectively.\cite{Ito97,Chou97,Kastner99,Eckert98,Yamada95} Below
$T_{N,I}$ a small spontaneous magnetization, M$_0$, within the basal plane of
the tetragonal lattice has been
reported.  The corresponding magnetic structure has been analysed in Refs.\ 
\onlinecite{Ito97,Chou97,Kastner99,Eckert98,Yamada95,Mueller99}. 

The aim of this paper is to develop the superexchange theory for the
orthorhombic compound
Ba$_3$Cu$_2$O$_4$Cl$_2$ (space group Pmma) with a Cu-O-Cu bond geometry rather
unusual in the 
cuprate family, but having also two crystallographically different types of Cu 
sites. \cite{Kipka2-76} Actually, in this compound the edge-sharing CuO$_4$
plaquettes are not 
aligned in the same plane, but form folded chains. The chain axis is parallel
to the orthorhombic $a$--axis. 
The unit cell with two sorts, A and B, of
crystallographically non-equivalent Cu sites is depicted in
Fig.~\ref{cell}. Due to the  
typical Cu$^{2+}$ (3$d^9$) states the compound is an insulator. 
It was found 
to behave like a classical antiferromagnet with 
a N\'eel temperature $T_N$ of 
about 20~K. \cite{Eckert98} Below $T_N$ and for applied magnetic fields H$_a$
parallel to the 
$a$--axis this compound shows a spin-flop transition 
at $\mu_0$H$_a$ $\approx$ 2.6T. \cite{Eckert98} Thus, 
the 'easy axis' of the antiferromagnetically ordered moments turns 
out to be the $a$--axis. Above $T_N$ the susceptibility 
follows a normal Curie-like behavior. 
From the Curie constant an effective paramagnetic moment of about 2 $\mu_B$ is 
derived, which is typical for Cu$^{2+}$ in the 3$d^9$ state. 
No weak ferromagnetism has been
observed. \cite{Eckert98} Preliminary group theoretical analysis showed 
\cite{Mueller99} that 
in Ba$_3$Cu$_2$O$_4$Cl$_2$ weak ferromagnetism is forbidden while weak
antiferromagnetism cannot be excluded. This symmetry analysis is extended
below in close relation with the microscopical consideration based on the
superexchange theory which yields Dzyaloshinskii-Moriya
\cite{Dzyal58,Moriya63} type interactions. 

The organisation of the paper is the following. In the next Section the
results of the bandstructure calculation in the local density approximation
(LDA) for Ba$_3$Cu$_2$O$_4$Cl$_2$ will be briefly presented. The basic set of
copper and oxygen orbitals responsible for superexchange processes will be
fixed together with the estimates for the corresponding electronic transfer
integrals and the orbital crystal field splitting parameters. At the next
step, to provide a background for the superexchange theory the underlying
multiorbital Hubbard model is formulated with spin-orbit coupling on copper
ions involved. The perturbation expansion of the multi-orbital Hubbard model
is used in Section III to derive an effective spin Hamiltonian for the
nearest-neighbour Cu-Cu magnetic interactions. A mean-field
analysis of the model derived is also presented and a prediction on the
ground-state magnetic structure in Ba$_2$Cu$_2$O$_4$Cl$_2$ is
formulated. A complementary group theoretical analysis is presented in Section 
IV. Concluding remarks can be found in Section V.

\section{Band structure and the underlying multi-orbital Hubbard model}

The bandstructure of Ba$_3$Cu$_2$O$_4$Cl$_2$ was calculated within the
local density approximation (LDA) using a recently developed
full-potential nonorthogonal local-orbital minimum-basis scheme
(FPLO).\cite{koepernik99} Here, the Cu(3$d$, 4$s$, 4$p$), Ba(5$s$,
5$p$, 5$d$, 6$s$, 6$p$), Cl(3$s$, 3$p$, 3$d$), and O(2$s$, 2$p$, 3$d$)
states were treated as valence states and the lower lying states as
core states.  For our purpose to extract tight-binding parameters it
is sufficient to perform a non-spinpolarized calculation.  The
non-magnetic solution shows metallic behavior with four bands crossing
the Fermi surface, corresponding to the four copper atoms per unit
cell. Hereafter the notations $B_1$, $A_1$, $B_2$ and $A_2$ will be used to
specify the different Cu sites in a unit cell.  

Due to the folding of the CuO$_2$ chains by nearly 90 degrees 
the bandwidth of the 
half-filled antibonding bands is relatively narrow ($\approx$ 0.5 eV)
and only half as large as in the planar (unfolded) edge-shared chains
such as in Li$_2$CuO$_2$ or CuGeO$_3$. \cite{r99} The orbital analysis for the
four 
bands at Fermi level shows that these bands are built up mainly from the
Cu$_A$--3$d_{yz}$ and the Cu$_B$--3$d_{x^2-y^2}$ orbitals, respectively,
with an admixture of all three O--2$p_x$, --2$p_y$, --2$p_z$ orbitals (the
global axes  
$x$, $y$, and $z$ are parallel to the crystallographic $a$--, $b$--  and 
$c$--axes, respectively). The other Cu--3$d$ orbitals give rise to a band
structure (16 bands altogether) which spans the range of binding energies from 
2 eV up to 4 eV below the Fermi level. The band states of predominantly O 
2$p_x$, 2$p_y$, 2$p_z$ character fall into the range of binding energies from 
$\sim$ 2 eV up to $\sim$ 6 eV. 

The indirect coupling of the Cu--3$d$ orbitals via the intermediate O--2$p$
orbitals can be described by a set of effective hopping matrix
elements. Considering first the top-most four bands formed by the active,
i.e.\ spin carrying, Cu$_A$--3$d_{yz}$ and Cu$_B$--3$d_{x^2-y^2}$ orbitals we
obtained the following results by fitting those bands to a 
four-band tight-binding (TB) model. 
\cite{rosner00}
The intra-chain hopping between the neighboring Cu$_A$ and
Cu$_B$ sites, $t_{AB}^x \simeq$ 65 meV, is of the same order as the
inter-chain diagonal Cu$_{B_1}$--Cu$_{B_1}$ hopping, 
$t_{BB}^{2xy} \simeq$~--~60~meV. 
Considering the hierarchy of
transfer processes, the 
next terms were found to be $t_{AA}^y \simeq t_{BB}^y \simeq$ 30 meV which
correspond to the inter-chain nearest neighbor hoppings along the
$y$--direction. 
The inter-layer hopping parameters along $z$--direction are very weak. 
Therefore, the seemingly quasi-one 
dimensional compound Ba$_3$Cu$_2$O$_4$Cl$_2$ has to be classified, in a first
approximation, as a 
two dimensional  compound with spatially anisotropic interactions within a
layer and weak 
couplings between the layers (which are necessary to explain the  
3D magnetism and the experimentally found finite 
N{\'e}el temperature). That is true also for the exchange couplings of the
spins related to the Cu$_A$--3$d_{xz}$ and Cu$_B$--3$d_{x^2-y^2}$ basic
orbitals. Below we will consider in great detail the microscopic origin of the 
dominant intra-chain nearest neighbor interactions that involve rather strong
magnetic anisotropies due to the non-trivial geometry of a particular chain. 
The remaining inter-chain couplings are expected to influence only the
spin isotropic part of the exchange interaction since they occur in a more
simple geometry. 

To develop the theory of superexchange, in the following the appropriate
underlying Hubbard model is formulated for Ba$_3$Cu$_2$O$_4$Cl$_2$. Due 
to the $S=1/2$ nature of the spins on the Cu$^{+2}$ sites, a 
single-ion anisotropy cannot occur. 
A magnetic anisotropy can only be obtained by taking into account
simultaneously both, the spin-orbit coupling and the splitting of the orbitals 
by crystalline fields. \cite{Coffey91} The spin-orbit coupling of the copper
ions is described by the $H_{\mbox{\small LS}}$ term in the underlying
electronic
Hamiltonian. This term and the kinetic one, $H_t$, are considered 
as the perturbation while the zero-order Hamiltonian 
$H_0 = H_0^{(d)} + H^{(p)}_0$ is
a sum of the on-site interactions at copper and oxygen atoms,
respectively.
Thus, the complete Hamiltonian is specified as follows
\begin{eqnarray}
H^{(d)}_0&=&\sum_j
\left\{\sum_m\sum_\alpha\varepsilon^d_{jm}d^\dagger_{jm,\alpha}d_{jm,\alpha} +
 \sum_{m m'}
\sum_{\alpha\alpha'} U_d^{mm'}n^d_{jm\alpha}n^d_{jm'\alpha'}\right\},
\nonumber\\
H^{(p)}_0 &=& \sum_l
\left\{\sum_k\sum_\beta\varepsilon^p_{lk}p^\dagger_{lk,\beta}p_{lk,\beta
} +
 \sum_{k k'}
\sum_{\beta\beta'} U_p^{kk'}n^p_{lk\beta}n^p_{lk'\beta'}\right\},
\nonumber\\
H_t &=& \sum_{j,m,\alpha}\sum_{l\in j,k}
\left(t_{jm,lk}d^\dagger_{jm,\alpha}p_{lk,\alpha} + h.c.
\right\},
\nonumber\\
H_{\mbox{\small LS}} &=& \lambda\sum_j \vec L_j\vec S_j ,
\label{1} 
\end{eqnarray}
where the hole representation and the standard notation for 
cuprates are used. Here $j$, which is a composite index, denotes a cell number 
and the sort of the Cu site in the cell; $l \in j$ is used to denote an O site
neighboring to the $j$--site. The term $\varepsilon^d_{jm}$ is the
crystal field level for the $m$--th copper $d$--orbital at the $j$--th
site. 
The results of the band structure calculations \cite{rosner00} allow  to
estimate  
the crystal field splitting between the lowest 
$\varepsilon^d_{j0}$ ($=\varepsilon^d_{0}$)
and the excited ($m\not=0$) levels $\varepsilon^d_{jm}$ ($\simeq
\varepsilon^d_{m}$) as $\varepsilon^d_{m} - \varepsilon^d_{0} \approx
2$eV
$\equiv \varepsilon_d$. We neglect the difference of this splitting 
between the different excited $m$--states as well as between the 
nonequivalent Cu $j$--sites, which is of minor importance for the
present 
purposes. For the further calculations only
the copper on-site Coulomb integral, $U^{mm'}_d$, with $m'=m$ is
required,
which is assumed in analogy with other cuprates to be 
$U^{mm}_d = U_d \simeq 8\div 10$eV. 
The crystal field splitting between the different oxygen
$|p_{lk}\rangle$--orbitals ($k=x,y,z$) located on a lattice site $l$ is not
taken into account in the present consideration. 
The estimate for the "bare" charge transfer gap 
$\Delta_p = \varepsilon^p - \varepsilon^d_0 \approx 4$eV is deduced from
the 
band structure calculations and a complementary finite cluster
analysis. \cite{Siurak99} For the oxygen on-site Coulomb
integrals $U^{kk}_p = U_p$ and $U^{kk'}_p = U_p - 2J_p$ (for $k \not=
k'$),
the following  estimates, $U_p \simeq 4$eV and $J_p \simeq 0.2 \div
0.4$eV, standard for the cuprates (see Ref.\ \onlinecite{Yush99} and references
therein),
are assumed.

Next, we consider the kinetic part, $H_t$. According to Slater and Koster (see 
Ref.\ \onlinecite{Harrison}) one can express the hopping
amplitudes $t_{jm,lk}$ between the nearest neighbor copper $|d_{jm}\rangle$--
and oxygen $|p_{lk}\rangle$--orbitals as linear combinations of the two
parameters, $(pd\sigma)$ and $(pd\pi)$. By using
the approximation $(pd\pi) \simeq -1/2 (pd\sigma)
= - t^{\mbox{\small eff}}_{pd}/\sqrt{3}$ which is valid for transition metal
oxides, \cite{Mattheiss} one may write $t_{jm,lk}=\chi_{mk} t^{\mbox{\small
eff}}_{pd}$. In this expression, for given $|d_{jm}\rangle$ and
$|p_{lk}\rangle$, the factor $\chi_{mk}$ is a function of direction cosines of
the $(\vec{j}-\vec{l})$ vector. All these factors, of order unity, are
calculated by using the routine procedure. \cite{Harrison}
The only parameter remaining, $t^{\mbox{\small eff}}_{pd}$, is estimated
from the bandstructure calculations. For the special composition of CuO$_4$
plaquettes in Ba$_3$Cu$_2$O$_4$Cl$_2$ with only nearest neighbor Cu-O hopping
we obtained the following estimate $t^{\mbox{\small eff}}_{pd} \simeq 0.5$
eV by using the formula $t_{AB}^x=(t^{\mbox{\small eff}}_{pd})^2/\Delta_p$.   
That value is smaller than the corresponding nearest neighbor Cu-O hopping
terms in other cuprates \cite{Eskes93,Yush99,Mattheiss89,bacuocl} since the
effects of 
the different geometry, the neglect of direct O-O transfer and of the crystal
field splitting at oxygen sites are all condensed into one effective
parameter. The value $\lambda=0.1$ eV of the spin-orbit coupling, 
characteristic for other cuprates, is taken in the calculation.

\section {Spin Hamiltonian and the mean-field ground-state spin
configuration}

The derivation of the effective superexchange interactions using
perturbation theory in the multi-band Hubbard model,
has been presented already many times for planar cuprates. 
\cite{Coffey91,Koshibae93,Koshibae94} 
Some peculiarities of the present derivation are due to the fact that the 
CuO$_2$-chains in Ba$_3$Cu$_2$O$_4$Cl$_2$ are folded. The unit cell of this
structure contains four different Cu ions which are denoted in Fig.\ 2 as
B$_1$, A$_1$, B$_2$ and A$_2$. Therefore, there is a sequence of four 
different 
magnetic bonds $(ij)$ = $(B_1,A_1)$, $(A_1,B_2)$, $(B_2,A_2)$ and
$(A_2,B'_1)$, where the site $B'_1$ belongs to the neighboring unit
cell. We start by deriving the spin Hamiltonian for a particular
(AB)-- or (BA)--bond of this sequence, and then extend the consideration to
the entire 
structure. Two oxygen ions, $(l=L,R)$, which are common neighbors of
Cu$_A$ and Cu$_B$, mediate the superexchange interaction between
Cu$_A$ and Cu$_B$. By applying an appropriate unitary transformation
one can define molecular-type basic functions for the oxygen
single-hole states in the form $|p_n\rangle= (|p_{Lk}\rangle \pm
|p_{Rk}\rangle)/\sqrt{2}$, with $(n=1, \dots 6$ for $k=x,y,z)$,
that are used below.

At the first step of the expansion procedure the excited 
$|d_{jm}\rangle$--orbitals ($m \not= 0$) are taken into account by 
additional vector hybrydization  terms 
$\sim \vec C_{j,n} \cdot \vec \sigma_{\alpha\beta}$ connecting the
   ground state $|d_{j0}\rangle$--orbitals with different
$|p_n\rangle$--orbitals. Then, the original hopping processes 
$\sim t_{j0,n}$ and the
additional new ones can be dealed simultaneously within the same  
approximation. The corresponding effective kinetic term is 
\begin{eqnarray}
H^{(AB)}_1 = \sum_{j=A,B}\sum_{n}\sum_{\alpha\beta}
\left[t_{j0,n}\delta_{\alpha\beta} + \vec
C_{j,n}\cdot\vec\sigma_{\alpha\beta}\right] d^\dagger_{j0, \alpha} p_{n,
\beta} + h.c.
\label{2}
\end{eqnarray}
where $\vec\sigma_{\alpha\beta}$ represent the Pauli matrices and
\begin{eqnarray}
\vec C_{j,n} = -\frac{\lambda}{2\varepsilon_d}\sum_m \vec L_{j, 0m}
t_{jm, n}\
\ ;\qquad \vec C^*_{n,j} = \vec C_{j,n} \quad .
\label{3}
\end{eqnarray}
 Here, $\vec L_{j, 0m}$ are matrix elements of the orbital angular 
 momentum operator and
 $t_{jm,n}$ are the original  amplitudes for the hopping process 
 between the excited
 $m$--th crystal-field $d$--state and the $n$--th oxygen $p$--state.

 Finally, the $|p_n\rangle$--orbital states can be eliminated in the 
perturbation
 procedure and the corresponding fourth-order processes are summed up
into 
 the following spin Hamiltonian \cite{Koshibae93}
referring to one Cu$_A$-(O-O)-Cu$_B$ magnetic bond 
\begin{eqnarray}
H^{(AB)} = J_{AB} \vec S_{A}\vec S_{B} + \vec D_{AB} \cdot \left[\vec
S_{A}
\times \vec S_{B}\right] + \vec
S_{A}\stackrel{\rightleftharpoons}{\Omega}_{AB}\vec 
S_{B}
\label{4}
\end{eqnarray}
with the interaction constants
\begin{eqnarray}
J_{AB} &=& 4\sum_{nn'} g_{nn'}\left[ t_{A0,n} t_{n, B0} + \vec C_{A,n}
\cdot
\vec C_{n,B}\right] \cdot \left[ t_{B0, n'} t_{n', A0} + \vec C_{B,n'}
\cdot
\vec C_{n',A}\right]\nonumber , \\ 
\vec D_{AB} &=& 8\imath \sum_{nn'} g_{nn'} t_{A0,n} t_{n, B0} \left[\vec
C_{B,n'}
t_{n', A0} + t_{B0, n'} 
\vec C_{n', A}\right]\nonumber , \\
\Omega^{\mu\nu}_{AB} &=& \Gamma^{\mu\nu}_{AB} + \Gamma^{\nu\mu}_{AB} -
\delta_{\mu\nu}\left( \sum_\xi \Gamma^{\xi\xi}_{AB  }\right)\label{5}
,\\ 
\Gamma_{AB}^{\mu\nu} &=& 4\sum_{nn'} g_{nn'}\left[ C^\mu_{A,n} t_{n, B0}
+ t_{A0,n} 
C^\mu_{n,B}\right] \cdot \left[C^\nu_{B, n'} t_{n', A0} + t_{B0,n'}
C^\nu_{n',A}\right] . \nonumber
\end{eqnarray}
Here $g_{nn'}$ is given by
\begin{eqnarray}
g_{nn'} = \frac{1}{\Delta^2_p}\left[\frac{1}{U_d} + \frac{1}{2\Delta_p}
+
\frac{1}{2\Delta_p + U_{p}^{nn'}}\right]
\label{6}
\end{eqnarray}
where $U_p^{nn} = U_p$ and $U^{nn'}_p = U_p - 2J_p \quad (n' \not= n)$.
Further on the notation $g_{nn} = g$ will be used. 
One may note a 
weak difference between $g_{nn}$ ($n' = n$) and $g_{nn'}$ ($n' \not= n$). 
Taking into account the particular form of the $t_{j0,n}$ ($= t_{n,j0}$)
and
$\vec C_{j,n}$ ($= \vec C^*_{n,j}$) transfer parameters, we find the
final
expressions for the non-zero interaction constants:
\begin{eqnarray}
J_{AB} &=& 4g b^2_{AB} - \delta J_{AB}\nonumber ,\\
D^y_{AB} &=& 8g b_{AB} c_{AB}\label{7} ,\\
\Gamma^{yy}_{AB} &=& 4g c^2_{AB} - \delta\Gamma^{yy}_{AB} ,\nonumber
\end{eqnarray}
where
\begin{eqnarray}
b_{AB} &=& \left(t^{\mbox{eff}}_{pd}\right)^2 \left(1 -
\cos\phi_{AB}\right) ,\nonumber\\
c_{AB} &=&
\frac{1}{\sqrt{3}}\left(\frac{\lambda}{\varepsilon_d}\right)\left(t^{\mbox{eff}}_{pd}\right)^2
\sin\phi_{AB} , \nonumber\\
\delta J_{AB} &\approx& \left(\frac{\lambda}{\varepsilon_d}\right)^2 4g
b^2_{AB}\label{8} ,\\
\delta \Gamma^{yy}_{AB} &\approx& \left(\frac{J_p}{2\Delta_p +
U_p}\right) 4g c^2_{AB} .\nonumber
\end{eqnarray}
It is worth emphasizing at this stage that the vector $\vec D_{AB}$ of
the
Dzyaloshinskii-Moriya interaction is directed along $\vec y$--axis
(i.e. $D^x_{AB} = D^z_{AB} = 0$) and the symmetric anisotropy is
described 
by only one non-zero parameter $\Gamma^{yy}_{AB}$. Note, that the
direction 
of $\vec D_{AB}$ is in agreement with the rules given by 
Moriya. \cite{Moriya63} The parameter $\phi_{AB}$ is characteristic 
for the directed $(AB)$--bond. More
generally, a particular parameter 
$\phi_{ij}$ has to be assigned to each bond in the entire chain-like structure 
in the following way.
Let us ascribe to the center of each $j$-th plaquette a local
$\vec{z_j}$--axis perpendicular to the plane of the plaquette. For instance,
in Fig.~\ref{Fig.2} the $\vec{z} (=\vec{z_i})$ and $\vec{z^\prime}
(=\vec{z_j})$ are related to 
$i=B_1$ and $j=A_1$, respectively. Then, $\phi_{ij}$ is the angle of the
rotation which transforms the local $\vec{z_i}$--axis into the
$\vec{z_j}$--axis. 
Therefore, one has
\begin{eqnarray}
\phi_{BA}^{11} = -\phi_{AB}^{12}= - \phi_{BA}^{22}= \phi_{AB'}^{21}
\simeq 84^\circ\label{a} ,
\end{eqnarray}
where the left (right) indices refer to the left (right) atom in the directed
bond.  
It can be directly checked that each magnetic bond is 
described by the same form of the spin Hamiltonian (\ref{4})--(\ref{8}). 
By using the relations (\ref{7}--\ref{a})
it can be easily seen that the parameters $J$ and
$\Gamma^{yy}$ are bond-independent while the component $D^y_{ij}$
changes the 
sign in accordance with (\ref{a}). 
The spin Hamiltonian for the entire chain-like magnetic system can be
written in the following form
\begin{eqnarray}
{\cal{H}} &=& \sum_{ij}{\cal{H}}^{(ij)}, \qquad {\cal{H}}^{(ij)} =
{\cal{H}}^{(ij)}_0 + \delta{\cal{H}}^{(ij)} ,\nonumber\\
{\cal{H}}^{(ij)}_0 &=& 4g\left(b^2\vec S_i \cdot \vec S_j +
2bc\xi_{ij}\vec
d \cdot\left[\vec S_i \times \vec S_j\right] + c^2\left\{2(\vec d\vec
S_i) (\vec
d\vec S_j) - \vec S_i \cdot \vec S_j\right\}\right) ,\nonumber\\
\delta{\cal{H}}^{(ij)} &=& -\delta J\vec S_i\vec S_j -
\delta\Gamma^{yy}\left\{2(\vec d\vec S_i)(\vec d\vec S_j) - \vec S_i\vec
S_j\right\} , \label{9}
\end{eqnarray}
where $\vec d$ is the unit vector along the global $\vec{b}$--axis.   
$\xi_{ij} = \mbox{sgn}\phi_{ij}$ determines the space pattern of  
the Dzyaloshinskii-Moriya vectors for the chain-like structure (see Eq.
(9)). 

Below the arguments given by Shekhtman, Entin-Wohlman and 
Aharony \cite{Shekhtman92} for the single-bond superexchange 
interactions are extended to the chain-like
magnetic system with this special pattern of the $\vec D_{ij}$ vectors. 
Actually, the spins in the lattice can be subdivided into two 
subsystems in such a way that the first 
(second) subsystem is
formed by the spins in $B$ ($A$) positions only. Let us now 
introduce the following redefinition of the spin variables
\begin{eqnarray}
\tilde{\vec S}_{B_1} &=& \vec S_{B_1}, \qquad \tilde{\vec S}_{B_2} =
\vec
S_{B_2} ,\nonumber\\
\tilde{\vec S}_{A_1} &=& \left(\vec d \cdot \vec S_{A_1}\right) \vec d +
\cos\theta\left[\vec S_{A_1} - \left(\vec d \cdot \vec
S_{A_1}\right)\vec d\right]
- k\sin\theta\left[\vec S_{A_1} \times \vec d\right] , \label{10}\\
\tilde{\vec S}_{A_2} &=& \left(\vec d \cdot \vec S_{A_2}\right) \vec d +
\cos\theta\left[\vec S_{A_2} - \left(\vec d \cdot \vec
S_{A_2}\right)\vec d\right]
+ k\sin\theta\left[\vec S_{A_2} \times \vec d\right] ,\nonumber
\end{eqnarray}
where $k = \pm 1$ and $k\tan\theta = -2bc/(b^2 - c^2)$. It should 
be noted, that the transformation (\ref{10}) corresponds to a 
rotation of the spins $\vec S_{A_1}$ ($\vec S_{A_2}$) by an angle 
$\theta$ ($-\theta$) with $\vec d$ as the rotation axis. The Hamiltonian
${\cal{H}}^{(ij)}_0$ is strictly transformed into an isotropic form 
\cite{Shekhtman92} while
concerning the remaining $\delta {\cal{H}}^{(ij)}$ an additional 
justification should be made. 
Actually, no terms higher than of second order in
$(\lambda/\varepsilon_d)$
have been kept up to now.
By noting that $\delta J, \delta\Gamma \sim (\lambda/\varepsilon_d)^2$
and in 
(\ref{10}) the angle $\theta \sim (\lambda/\varepsilon_d)$, the
consistent 
way is to keep the form of $\delta{\cal{H}}^{(ij)}$ but using the 
transformed spin variables. 
Finally the transformed Hamiltonian $\tilde{\cal{H}} =
\sum\tilde{\cal{H}}^{(ij)}$ takes the form
\begin{eqnarray}
\tilde{\cal{H}}^{(ij)} = \bar{J}\tilde{\vec S_i}\tilde{\vec S_j} -
\bar{\Gamma}\left\{2(\vec d\tilde{\vec S_i})(\vec d\tilde{\vec S_j}) -
\left(\tilde{\vec S_i} \tilde{\vec S_j}\right)\right\}\label{11}
\end{eqnarray}
with the renormalized constants  $\bar{J} = 4g(b^2 + c^2) - \delta J$,\
\ 
and $\bar{\Gamma} = \delta\Gamma^{yy} > 0$.

A mean field analysis of this Hamiltonian shows that the expected
classical
ground-state configuration is a collinear antiferromagnetic array of the
transformed spins (\ref{10}) in the magnetically easy $xz$--plane,
$\bar{\Gamma} > 0$.  
We emphasize that according to (\ref{11}) the staggered moment is not
confined to a particular direction in the 
easy $xz$--plane. This 'residual'
symmetry of the derived superexchange can be broken by additional 
interaction terms 
not included into the consideration up to now. Postponing a discussion
of 
possible sources for additional anisotropies let us now assume that the 
'residual'  symmetry is broken in such a way that the 
staggered magnetization is parallel to 
the $\vec x$--axis (i.e. crystallographic $a$--direction). 
It means that in the ground state the spins
$\tilde{\vec S}_{B_1}$ ($= \vec S_{B_1}$) and $\tilde{\vec
S}_{B_2}$ ($= \vec S_{B_2}$) are aligned in the $\vec x$--direction. 
In the mean-field approximation one  may  immediately write
\begin{eqnarray}
\tilde{\vec S}_{B_1} =-\tilde{\vec S}_{A_1} = \tilde{\vec S}_{B_2} =
-\tilde{\vec
S}_{A_2} = S \vec e_x
\label{12}
\end{eqnarray}
and performing the inverse transformation one obtains for the original
spins
\begin{eqnarray}
\vec S_{B_1} &=& \vec S_{B_2} = S \vec e_x , \nonumber\\
\vec S_{A_1} &=& -S\left[\cos\theta \cdot \vec e_x + k\sin\theta \cdot
\vec
e_z\right] , \label{13}\\
\vec S_{A_2} &=& -S\left[\cos\theta \cdot \vec e_x - k\sin\theta \cdot
\vec
e_z\right] . \nonumber
\end{eqnarray}
The corresponding picture for the double degenerate classical
ground-state spin 
configuration is given in Fig.~3. A weak transverse 
(along $\vec z$--axis) modulation in the second sublattice is imposed on
the strong antiferromagnetic correlations between the spins belonging to
different sublattices. This weak antiferromagnetic modulation 
is entirely due to the Dzyaloshinskii-Moriya interaction. 
The angle $\theta$ for this modulation is estimated to be  
\begin{eqnarray}
\theta \simeq \frac{|\vec D_{ij}|}{J} \simeq \frac{2c}{b} =
\frac{2}{\sqrt{3}}\left(\frac{\lambda}{\varepsilon_d}\right)
\left|\sin\phi_{ij}\right| \simeq 3^\circ .
\label{14}
\end{eqnarray}
with an absolute value of $|\vec D_{ij}|$ $\sim$ 0.25 meV corresponding 
to $J$ $\sim$ 5 meV.
It is worth mentioning that the magnetization measurements
\cite{Eckert98} 
for Ba$_3$Cu$_2$O$_4$Cl$_2$ have clearly shown that the
crystallographic $a$--direction ($\vec x$--axis in our notations) is
the preferred direction for the staggered moment below $T_{N}$, where
$T_{N} \simeq 20$K is the temperature of the three dimensional
antiferromagnetic ordering in Ba$_3$Cu$_2$O$_4$Cl$_2$.  According to
this analysis it is expected that the ground state in this compound is
not a simple collinear antiferromagnetic configuration, but involves
also a weak antiferromagnetic superstructure due to the
Dzyaloshinskii-Moriya interaction. 

To our knowledge, in cuprates only weak ferromagnetism (WFM) was reported up
to now. The corresponding spin canting angle, which measures the deviation 
from collinearity, is rather small, $\theta_{\mbox{\small WFM}} \simeq
0.05^\circ$ (compare for instance Refs.\
\onlinecite{Thio88,Coffey91,Koshibae93,Koshibae94}). In the present case the
spin 
canting angle (\ref{14}), i.e.\ $\theta \simeq 3^\circ$, is more than
an order of magnitude larger. The reason for this difference is the
following. Although the spin-orbit coupling $(\lambda/\varepsilon_d)\sim 0.1$
is nearly the same in both cases, the geometrical factor is much larger in the 
present case as $\left|\sin\phi_{ij}\right| \simeq 1$ due to the 
strong folding of the chains ($\left|\phi_{ij}\right| \simeq 84^\circ$). 

It is necessary to discuss the possible influence of 
transverse inter-chain magnetic interactions to show the validity of the
derived spin anisotropy. We assume that the
dominant interaction between two neighboring spins belonging to different
chains in a layer is the isotropic superexchange with an AFM interaction
constant $J_{\perp}$. By using the results of the bandstructure calculations
(Section 
II) we found the estimate $J_{\perp}/J \sim 0.25$, where $J$ is the
intra-chain exchange constant. Due to the presence of the center of inversion
for the transverse Cu-Cu magnetic bond a Dzyaloshinskii-Moriya term cannot
occur and the symmetric anisotropy is expected to be very weak. 
The bandstructure results indicate also a weak inter-layer exchange
$J_z \ll J_{\perp},J$ and more distant inter-chain Cu-Cu exchange interactions 
(for instance to third neighbors in the plane of Cu$_B$). However, the spin
anisotropy of the latter interaction is expected to be small due to the planar 
geometry of the corresponding exchange path. 

\section{Symmetry and Spatial Ordering of the Dzyaloshinskii-Moriya
Vectors}

In order to derive the spatial order of the Dzyaloshinskii-Moriya 
vectors by symmetry considerations, at first the symmetry analysis of 
the antiferromagnetic states in Ba$_3$Cu$_2$O$_4$Cl$_2$ (following Refs.
\onlinecite{Turov65,Borovik68,Bertaut63}), presented elsewhere 
\cite{Mueller99}, will be summarized. It is assumed that the lattice 
constant $a$ of the magnetic unit cell and that of the
crystallographic unit cell are identical. As has been pointed out in
Ref. \onlinecite{Bertaut63} it is sufficient to consider the
independent symmetry operations. For the space group Pmma besides the
identity there are only three independent symmetry elements.

A very convenient choice for these elements is: rotation axis C$_{2y}$
through the point (x$_0$=0,z$_0$=1/2), rotation axis C$_{2z}$ through
the point (x$_0$=1/4,y$_0$=0) and inversion I located at
(x$_0$=0,y$_0$=0,z$_0$=1/2).  As these elements map a given chain onto
itself no assumptions concerning the relative orientation of moments
in different chains are necessary.  The magnetic moments reside on the
2c--sites Cu$_{A1}$, Cu$_{A2}$ and Cu$_{A'1}$, and on the 2e--sites
Cu$_{B1}$, Cu$_{B2}$ and Cu$_{B'1}$ (see Fig.~\ref{Fig.4}, in which
also the mentioned symmetry elements are sketched). Obviously the ion
Cu$_{A1}$ (Cu$_{B1}$) is equivalent to Cu$_{A'1}$
(Cu$_{B'1}$). Because the sites of the 2c-- and 2e--sublattices are
crystallographically not equivalent, the magnetic order has to be
characterized by two antiferromagnetic (AFM) vectors
\begin{eqnarray}
\vec L_{A} = \vec M_{A1} - \vec M_{A2}  \hspace{1cm}      \vec L_{B} =
\vec M_{B1} - \vec M_{B2}				
\label{15}
\end{eqnarray}
and two sublattice magnetizations 
\begin{eqnarray}
\vec m_{A} = (\vec M_{A1} + \vec M_{A2})/2  \hspace{1cm}      \vec m_{B}
= (\vec M_{B1} + \vec M_{B2})/2 
\label{16}
\end{eqnarray}
where $\vec M_{Ai}$ ($\vec M_{Bj}$) are the magnetic moments of the
Cu$_{Ai}$ (Cu$_{Bj}$) ions at the sites A$_{i}$ (B$_{j}$).  The
interchange of the moments $\vec M_i$ $\leftrightarrow$ $\vec M_j$
generated by the symmetry operations is given in the second and third
column of Table I.  As only one ordering temperature is found, both
AFM vectors have to transform according to the same magnetic group,
compatible with the crystallographic space group Pmma. The
transformation behavior for each component $L_{A,x}$, $L_{A,y}$
etc. and for the sublattice magnetizations $\vec m_{A}$ and $\vec
m_{B}$ is described in Table I by two numbers characterizing two steps
of the corresponding transformation: the transformation of $\vec
L_{A}$, $\vec L_{B}$ etc. due to $\vec M_i$ $\leftrightarrow$ $\vec
M_j$ (first number) and the remaining part of the transformation
(rotation, inversion; second). The entry +1 means no change of the
considered vector component, -1 means a change of sign. In the line
below the resulting magnetic symmetry element is given.  As I$\,\vec
L_{A}$ = $\vec L_{A}$, but RI$\,\vec L_{B}$ = $\vec L_{B}$ the AFM
vectors $\vec L_{2c}$ and $\vec L_{2e}$ cannot simultaneously become
different from zero. As each component of $\vec m_{A}$ and $\vec
m_{B}$ transforms according to the same magnetic group, the only
possibility for antiferromagnetic ordering is $L_x$ = $m_{A,x}$ -
$m_{B,x}$ $\neq$ 0, $L_y$ = $m_{A,y}$ - $m_{B,y}$ $\neq$ 0 {\bf or}
$L_z$ = $m_{A,z}$ - $m_{B,z}$ $\neq$ 0.  This means, ferromagnetic
order is predicted for each of the two crystallographic Cu
sublattices. The moments of the crystallographically different
sublattices are antiparallel to each other.  As different components
of $\vec m_A$ and $\vec m_B$ belong to different magnetic groups
(Table I) (i) only one component of the AFM vector $\vec L$ may be
different from zero, i.e. $\vec L$ is parallel to one of the
crystallographic axes, and (ii) weak ferromagnetism is excluded,
contrary to the case of Ba$_2$Cu$_3$O$_4$Cl$_2$. This is in agreement
with the experiments.  The symmetry analysis cannot predict the
direction of the AFM vector. Experiments show $L_x$ $\neq$
0.\cite{Eckert98} For that case, independent magnetic symmetry
elements are E, RC$_{2z}$, RC$_{2y}$ and I. Interestingly, Table I
shows, that for that case a small $L_{A,z}$ $\neq$ 0 i.e. weak
antiferromagnetism \cite{Turov65} can additionally occur.

As shown in Sect. III, within the unit cell this weak 
antiferromagnetism is described by a sum of Dzyaloshinskii-Moriya terms
\begin{eqnarray}
H_{WAFM}&=&   D_{BA}^{11} \cdot (M_{B1,x} M_{A1,z} - M_{B1,z}
M_{A1,x})\nonumber\\
        &+ &  D_{AB}^{12} \cdot (M_{A1,x} M_{B2,z} - M_{A1,z}
M_{B2,x})\nonumber\\
        &+ &  D_{BA}^{22} \cdot (M_{B2,x} M_{A2,z} - M_{B2,z}
M_{A2,x})\nonumber\\
        &+ &  D_{AB'}^{21} \cdot (M_{A2,x} M_{B'1,z} - M_{A2,z}
M_{B'1,x}) .
\label{17}
\end{eqnarray}  
This interaction has to be invariant with respect to the symmetry 
operations RC$_{2z}$, RC$_{2y}$ and I. Using the transformation 
properties of the moments according to Table I, the equivalence of sites
differing by a lattice translation  along the $a$--axis, results in 
\begin{eqnarray}
D_{BA}^{11} = -D_{AB}^{12} = -D_{BA}^{22} =  D_{AB'}^{21} .
\label{18}
\end{eqnarray}  
As to be expected, the spatial order of the local
Dzyaloshinskii-Moriya vectors determined from the symmetry of the
magnetic unit cell is the same as that derived by analyzing the
relations between the different bonds in the previous
Section. Moreover the symmetry analysis together with the experimental
results shows, that in the transformation described in Sect.~III
(canting of the moments) the Cu$_{Bi}$ moments have to be fixed, as
$L_{B}$ has to remain zero for an AFM state with moments parallel to
the $a$--axis.

\section{Conclusions}

The superexchange interaction between two neighboring but
crystallographically non-equivalent Cu ions in Ba$_3$Cu$_2$O$_4$Cl$_2$
was analyzed within a multi-orbital Hubbard model. This superexchange
interaction was expressed by an effective spin-spin Hamiltonian. As
there is no center of inversion for the considered Cu-(O-O)-Cu entity,
this interaction involves the Dzyaloshinskii-Moriya term, which refers
to the corresponding bond. By means of an analysis of the geometric
relations between neighboring bonds (plaquettes) the spatial order of
these locally defined Dzyaloshinskii-Moriya vectors was
determined. This spatial order was also derived by a symmetry analysis
taking into account the experimentally found 'easy axis' of the
staggered magnetization. These considerations revealed, that spin
canting occurs in the subsystem of the Cu$_A$ magnetic moments
only. Within the presented microscopic model, for the magnetic moments
all directions within the $a$-$c$--plane are equivalent. The
anisotropy, leading to the experimentally observed spin-flop
transition for applied fields parallel to the $a$--axis could not yet be  
explained. Further investigations have to be done to find the
microscopic origin for additional anisotropies, which break the
easy-plane symmetry of the superexchange model and explain the
experimentally observed easy-axis behavior. With this respect, two
main contributions have been ignored in the considered Hamiltonian
(\ref{1}). First of all, the theory of superexchange can be developed
at a more sophisticated level by adopting more details of the
electronic structure of the real compound.  For instance, the form of
the symmetric anisotropy tensor is rather sensitive to an actual
crystal-field splitting of $p$--orbitals on oxygen ions. This
splitting should be taken into account into the theoretical scheme. A
second reason leading to a breakdown of the easy-plane symmetry
involves the direct-exchange contribution
\cite{Shekhtman93} to the symmetric anisotropy term. This is due to the 
exchange part of the two-site Coulomb multi-orbital $d-d$ correlations
which should be also incorporated into the Hamiltonian (\ref{1}).
However, for such calculations reliable quantitative estimates for the
parameters are required (i.e. the crystal-field states of the O--ions
and the two-site exchange Coulomb integrals of Cu--ions).  Since the
direct exchange contribution can be hardly obtained within the LDA
band structure analysis, we have to leave this problem for future
investigations.  From an experimentally point of view, the detection
of the discussed weak antiferromagnetism is a challenging task.

\vspace*{1cm}

{\bf Acknowledgement}

The authors thank the INTAS organization (INTAS-97-11066) and the
Graduiertenkolleg at the TU Dresden (DFG) for support. We are grateful to
M.~L\"owenhaupt and A.~Krey\ss ig for fruitful discussions.

\newpage

\begin{figure}
\caption{Two unit cells of Ba$_3$Cu$_2$O$_4$Cl$_2$, space group Pmma. 
The Cu-O$_4$ plaquettes form folded chains with their axes parallel to 
the $a$--axis. To better demonstrate one of these CuO$_2$--chains, 
the origin in the figure was taken at (0,1/2,0). The crystal 
structure as well as the chains contain two types of 
copper sites, Cu$_A$  and Cu$_B$. }  
\label{cell}
\end{figure}

\begin{figure}
\caption{ Spatial ordering of the Dzyaloshinskii-Moriya 
vectors  $\vec D_{BA}^{ij}$ etc. related to a pair of Cu atoms located
at 
B$_i$ and A$_j$. (The first indices refer to the left atom in the
pair). 
The $z$-- ($z'$--) axes are locally defined and are perpendicular to the 
corresponding plaquette. The dashed line is the local $z$-- ($z'$--) axis 
of the plaquette left to that under consideration and is used for 
the definition of the angles $\phi_{BA}^{ij}$: The rotation of the 
corresponding dashed line onto the local axis results in the rotation 
angle $\phi_{BA}^{ij}$, the sign of which defines the direction of 
the local Dzyaloshinskii-Moriya vectors $\vec D_{BA}^{ij}$.  
The crystallographic axes $a$, $b$, and $c$ form the 
global coordinate system.}
\label{Fig.2}
\end{figure}

\begin{figure}
\caption{ Within a classical picture (mean field approximation) 
the Dzyaloshinskii-Moriya interaction 
between the magnetic moments (arrows) of 
the Cu$_A$ and Cu$_B$ atoms results in a small canting of 
the Cu$_A$ moments (canting angle $\pm  \theta$). The ground 
state is two-fold degenerated (a,b) and reveals weak antiferromagnetism.
The two states differ by the sign of the canting angle and the direction
of the of the additional antiferromagnetic 
vector $L_{A,z}$ = $M_{A1,z}$ - $M_{A2,z}$ of the Cu$_A$ sublattice.
x and z are global axes equivalent to the crystallographic 
$a$-- and $c$--axes. }  
\label{Fig.3}
\end{figure}

\begin{figure}
\caption{Symmetry elements of a Cu$_A$-Cu$_B$--chain 
in Ba$_3$Cu$_2$O$_4$Cl$_2$. The magnetic moments are located at the 
2c--sites (A$_1$, A$_2$ and A'$_1$) and at the 
2e--sites (B$_1$, B$_2$ and B'$_1$). The symmetry elements mapping the 
chain onto itself are: axes C$_{2z}$ going through the 
B-sites, axes C$_{2y}$ (sketched by small ellipses) perpendicular to 
the $x$-$z$--plane and going through the A--sites, and inversion centers  
located at the sites A$_i$. }  
\label{Fig.4}
\end{figure}

\newpage

\begin{table}
\caption{The first column lists the independent symmetry elements  
of the space group Pmma, here chosen to be E, C$_{2z}$, C$_{2y}$ 
and I. The second and third columns give the magnetic
moments $\vec M_{Ai}$ and $\vec M_{Bj}$ (i, j = 1, 2) generated 
from $\vec M_{A1}$ and $\vec M_{B1}$ by applying 
these symmetry elements. In the remaining columns for each 
component of the antiferromagnetic vectors 
$\vec L_{A}$ = $\vec M_{A1} - \vec M_{A2}$, 
$\vec L_{B}$ = $\vec M_{B1} - \vec M_{B2}$, 
and the two sublattice magnetizations 
$\vec m_{A}$ = $(\vec M_{A1} + \vec M_{A2})/2$ and 
$\vec m_{B}$ = $(\vec M_{B1} + \vec M_{B2})/2$ 
two numbers are given: the first takes into account the 
interchange $\vec M_{A1}$ $\leftrightarrow$ $\vec M_{A2}$, 
$\vec M_{B1}$ $\leftrightarrow$ $\vec M_{B2}$ (1 $\leftrightarrow$ no 
change of the vector components,  -1 $\leftrightarrow$ vector components
change sign). The second  +1 or -1 describes the remaining part of the 
considered transfomation (rotation, inversion). The resulting magnetic 
group elements are given in the line below  
these two numbers (R - time reversal).}
\label{TabI}
\end{table}

\begin{tabular}{cccccccccccccccc}
\hline
\hline
 E        &\qquad & $\vec M_{A1}$& $\vec M_{B1}$ &\qquad  & $L_{A,x}$ &
$L_{A,y}$ & $L_{A,z}$  &\qquad & $L_{B,x}$ & $L_{B,y}$ &  $L_{B,z}$
&\qquad &  $m_{A,x}$ & $m_{A,y}$ & $m_{A,z}$ \\       
          &\qquad &   (2c-site)  &  (2e-site)    &\qquad  &           &
&            &\qquad &           &           &             &\qquad &
$m_{B,x}$ & $m_{B,y}$ & $m_{B,z}$ \\ \hline      
 C$_{2z}$ &\qquad &$\vec M_{A2}$ & $\vec M_{B1}$ &\qquad  &  -1,-1    &
-1,-1   &   -1,1     &\qquad &  1,-1     &   1,-1    &    1,1
&\qquad &   1,-1     &    1,-1   &   1,1     \\       
          &\qquad &              &               &\qquad  & C$_{2z}$  &
C$_{2z}$  & RC$_{2z}$  &\qquad & RC$_{2z}$ & RC$_{2z}$ &  C$_{2z}$
&\qquad &  RC$_{2z}$ & RC$_{2z}$ &  C$_{2z}$ \\ \hline      
 C$_{2y}$ &\qquad & $\vec M_{A1}$& $\vec M_{B2}$ &\qquad  & 1,-1      &
1,1     &    1,-1    &\qquad &  -1,-1    &   -1,1    &   -1,-1
&\qquad &   1,-1     &    1,1    &   1,-1    \\       
          &\qquad &              &               &\qquad  & RC$_{2y}$ &
C$_{2y}$  & RC$_{2y}$  &\qquad &  C$_{2y}$ & RC$_{2y}$ &   C$_{2y}$
&\qquad &  RC$_{2y}$ & C$_{2y}$  & RC$_{2y}$ \\ \hline      
 I        &\qquad &$\vec M_{A1}$ & $\vec M_{B2}$ &\qquad  &  1,1      &
1,1     &    1,1     &\qquad &  -1,1     &   -1,1    &    -1,1
&\qquad &   1,1      &    1,1    &   1,1     \\      
          &\qquad &              &               &\qquad  &   I       &
I      &     I      &\qquad &    RI     &    RI     &     RI
&\qquad &     I      &     I     &    I      \\ \hline        
\hline
\end{tabular}

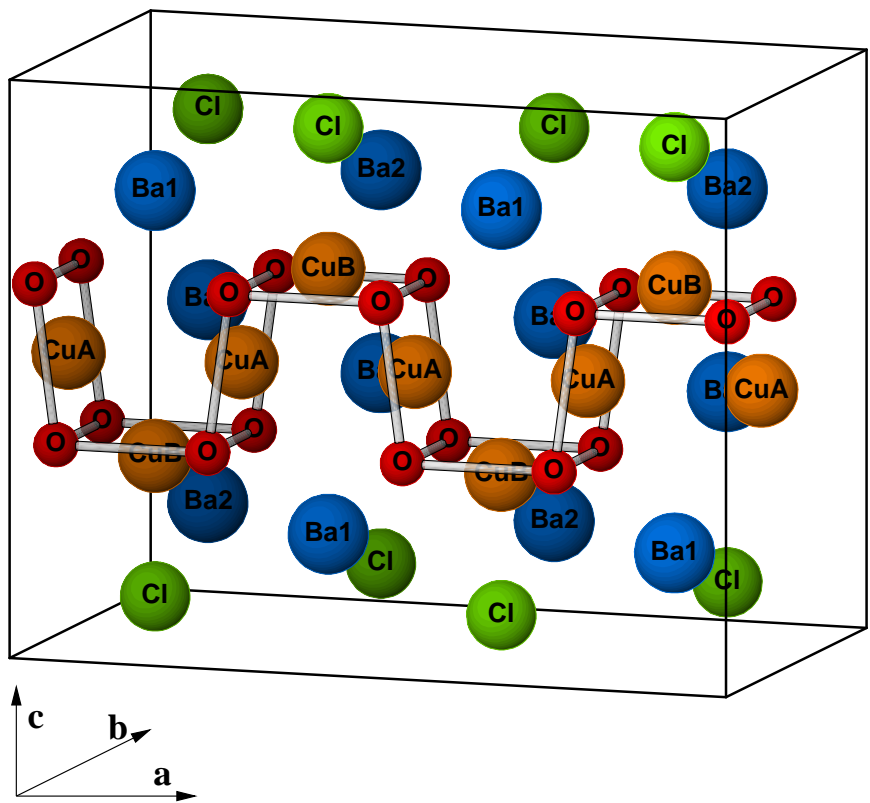
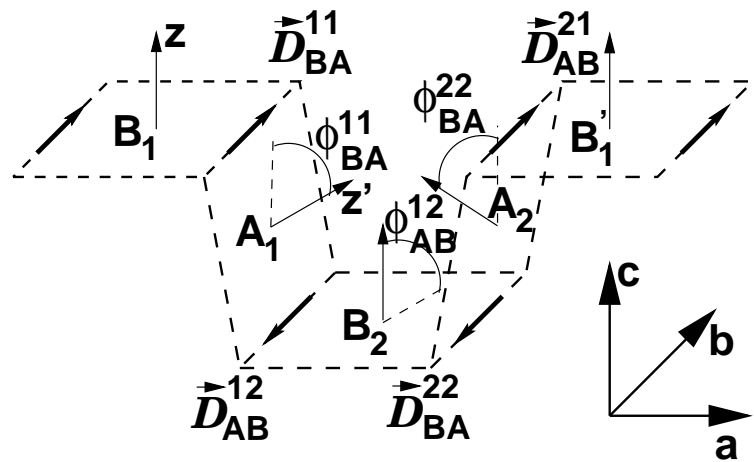
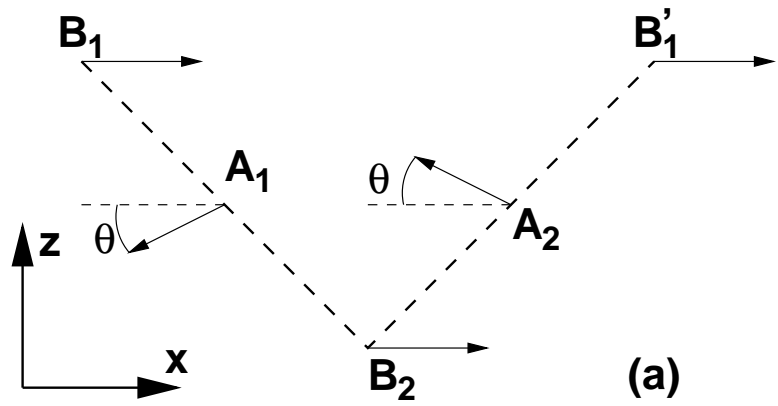
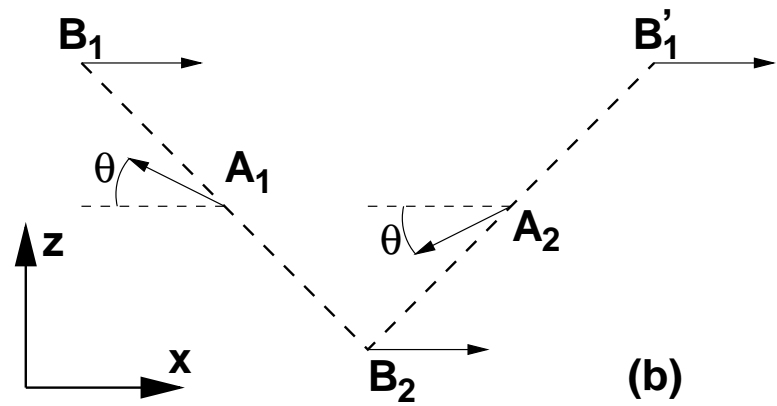
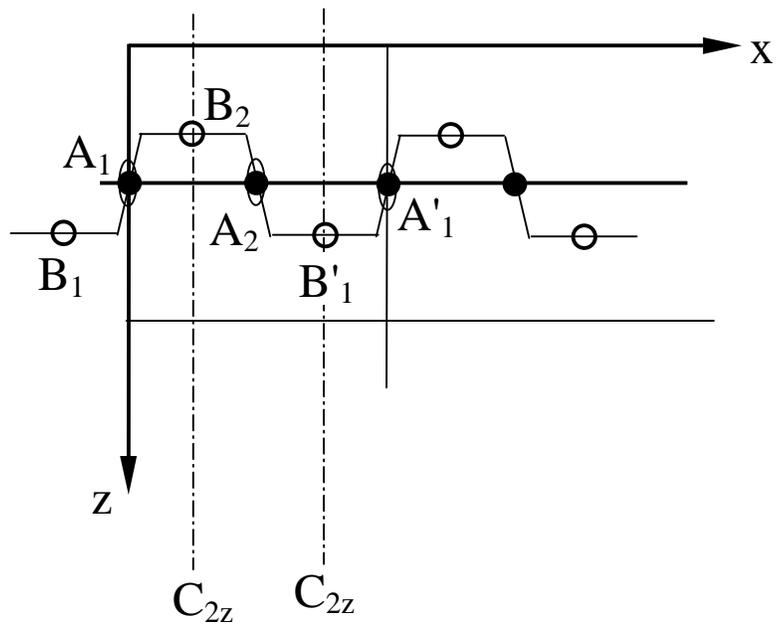
\end{document}